\documentclass[aps,nofootinbib,floatfix,showpacs,twocolumn]{revtex4}
\usepackage{mathrsfs}
\usepackage{graphicx}
\usepackage{amsmath}
\usepackage{amsfonts}
\usepackage{amssymb}
\usepackage{color}

\usepackage{epsfig}
\usepackage{CJK}
\usepackage{graphicx}
\usepackage{epsfig}
\usepackage{eepic}
\usepackage{bbm}
\usepackage{dcolumn}
\usepackage{bm}
\usepackage{ulem}
\usepackage{slashbox}
\usepackage{multirow}
\usepackage{slashed}

\newcommand{\omits}[1]{}

\def\bc{\begin{center}}

\def\ec{\end{center}}
\def\be{\begin{eqnarray}}
\def\ee{\end{eqnarray}}

\definecolor{dyellow}{rgb}{1.,0.8,.0}
\definecolor{myblue}{rgb}{.1,.1,.7}
\definecolor{dcyan}{rgb}{.0,.6,.6}
\definecolor{cyan}{rgb}{0.4,1.0,1.0}
\definecolor{dmagenta}{rgb}{0.6,0.0,0.6}
\definecolor{brown}{rgb}{0.6,0.2,0.}
\definecolor{darkblue}{rgb}{.0,.0,0.5}
\definecolor{darkred}{rgb}{0.75,0.0,0.0}
\definecolor{orange}{rgb}{1.,.6,.0}
\definecolor{dorange}{rgb}{0.8,.4,.0}
\definecolor{green}{rgb}{0.0,1.0,0.0}
\definecolor{darkgreen}{rgb}{0.0,0.6,0.0}
\definecolor{purple}{rgb}{.4,.0,.4}
\definecolor{lightgrey}{rgb}{0.7, 0.7, 0.7}
\definecolor{grey}{rgb}{0.4, 0.4, 0.4}

\newcommand{\nc}{\newcommand}
\nc{\rnc}{\renewcommand} \nc{\ket}[1]{\left | \, #1 \right \rangle}
\nc{\bra}[1]{\left \langle #1 \, \right |}
\nc{\ua}{\uparrow} \nc{\da}{\downarrow}

\nc{\braket}[2]{\langle\, #1\,|\,#2\,\rangle}
\nc{\half}{\frac{1}{2}}

\nc{\prj}{\mathcal{P}} \nc{\hilb}{\mathcal{H}}
\nc{\pth}{\mathcal{C}} \nc{\inprod}[2]{\braket{#1}{#2}}
\nc{\upket}{\ket{\uparrow}} \nc{\downket}{\ket{\downarrow}}
\nc{\upbra}{\bra{\uparrow}} \nc{\downbra}{\bra{\downarrow}}

\begin{document}


\title{Thread/State correspondence: the qubit threads model of holographic gravity}

\author{Yi-Yu Lin$^{1,2}$} \email{yiyu@bimsa.cn}
\author{Jie-Chen Jin$^3$} \email{jinjch5@mail2.sysu.edu.cn}

\affiliation{${}^1$Beijing Institute of Mathematical Sciences and Applications (BIMSA),
	Beijing, 101408, China}
\affiliation{${}^2$Yau Mathematical Sciences Center (YMSC), Tsinghua University,
	Beijing, 100084, China}
\affiliation{${}^3$School of Physics and Astronomy, Sun Yat-Sen University, Guangzhou 510275, China}


\begin{abstract}
We construct a new toy model of the holographic principle, named as holographic qubit threads model, which is an enlightening step towards the issue of spacetime emergence (``it from qubit''). More specifically, we propose for the first time that each bit thread in a locking bit thread configuration is in a ``qubit'' state, i.e., the quantum superposition state of two orthogonal states. Using this thread/state correspondence, we can construct the explicit expressions for the SS states corresponding to a set of bulk extremal surfaces in the surface/state correspondence, and nicely characterize their entanglement structure. Then we use the locking bit thread configurations to construct the holographic qubit threads model, and we show that it is closely related to the holographic tensor networks, kinematic space, and the connectivity of spacetime.

\end{abstract}

\pacs{04.62.+v, 04.70.Dy, 12.20.-m}

\maketitle

\section{Introduction}

How to unify or reconcile quantum mechanics and general relativity is undoubtedly the most important problem in physics. Advances in recent decades suggest that we may have found an important and marvelous clue to the mystery between gravity and quantum mechanics. This clue is successively related to the concepts of the black hole area entropy~\cite{Bekenstein:1972tm,Bekenstein:1973ur,Bekenstein:1974ax,Bardeen:1973gs}, the holographic principle (especially AdS/CFT duality)~\cite{Maldacena:1997re,Gubser:1998bc,Witten:1998qj}, and the RT formula of the holographic entanglement entropy~\cite{Ryu:2006bv,Ryu:2006ef,Hubeny:2007xt}, which have built a bridge between quantum mechanics and general relativity. In particular, the RT formula shows that the entanglement entropy that characterizes quantum entanglement between different parts of a particular class of (i.e., ``holographic") quantum systems can be equivalently (i.e., ``dually") described by the area of an extremal surface in a corresponding curved spacetime~\cite{Ryu:2006bv,Ryu:2006ef,Hubeny:2007xt}. 

More recently, many enlightening ideas from other fields, such as condensed matter physics, quantum information theory, network flow optimization theory, etc., have entered and benefited the study of the holographic gravity. One of the most striking examples is that inspired by the tensor network method originally used in condensed matter physics as a numerical simulation tool to investigate the wave functions of quantum many-body systems, various holographic tensor network (TN) models have been constructed as toy models of the holographic duality~\cite{Vidal:2007hda,Vidal:2008zz,Swingle:2009bg,Swingle:2012wq, Pastawski:2015qua, Hayden:2016cfa, Chen:2021ipv, Bao:2018pvs,Bao:2019fpq,Lin:2020ufd, SinaiKunkolienkar:2016lgg,  Qi:2013caa, Sun:2019ycv, Haegeman:2011uy}. Furthermore, inspired by the continuous version of the holographic tensor network models,~\cite{Miyaji:2015fia,Miyaji:2015yva} proposed the so-called SS correspondence (surface/state correspondence) as a more specific mechanism of the holographic principle. The SS duality refers to the duality between a codimension two convex surface $\Sigma $ in the holographic bulk spacetime and a quantum state described by a density matrix $\rho (\Sigma )$, which is defined on the Hilbert space of the quantum theory dual to the Einstein's gravity. 

Another idea to further explore the profound connection between spacetime geometry and quantum entanglement is inspired by the optimization problem in network flow theory.~\cite{Freedman:2016zud,Cui:2018dyq,Headrick:2017ucz} developed the optimization theory of flows on manifolds, endowed with the name of ``bit threads", and proposed the concept of bit threads can equivalently formulate the RT formula. In this letter, by studying the connection between bit threads and SS duality, we propose a natural and novel physical property of bit threads, dubbed ``thread/state correspondence'', that is, in a so-called locking bit thread configuration~\cite{Headrick:2020gyq,Lin:2020yzf,Lin:2021hqs,Lin:2022aqf}, each bit thread is in a quantum superposition state of two orthogonal states. Since the flux of bit threads is directly related to the geometry of a holographic spacetime, while this thread-state property can cleverly characterize quantum entanglement as one will see, it can be expected to be a very useful advance in the study of the relationship between quantum entanglement and spacetime geometry. Note that because bit threads are visually intuitive, they have usually been interpreted implicitly as the bell pairs distilled from the boundary quantum systems, however, they are mostly used merely as a mathematical tool to study different aspects of the holographic principle~(see e.g.~\cite{Lin:2022aqf,Lin:2021hqs,Lin:2020yzf,Headrick:2020gyq,Agon:2021tia,Rolph:2021hgz,Hubeny:2018bri,Agon:2018lwq,Du:2019emy,Harper:2019lff,Agon:2019qgh,Agon:2020mvu,Pedraza:2021fgp,Pedraza:2021mkh,Harper:2018sdd,Shaghoulian:2022fop,Susskind:2021esx,Kudler-Flam:2019oru,Headrick:2022nbe,Harper:2021uuq,Harper:2022sky}). Our thread/state rules explicitly accentuate for the first time the implied meaning of the name ``bit threads'', namely, that these threads, which are used mathematically to recover the RT formula, can be physically assigned a meaning closely related to the concept of ``bits''.

Using thread/state rules, we can do (but expect to do more than) the following things: we can construct the explicit expressions for the SS states corresponding to a set of bulk extremal surfaces in the SS duality, and nicely characterizing their entanglement structure; use the locking bit thread configurations to construct a new toy model of the holographic principle and actually, we explicitly give the relationship between this model and the holographic tensor network models; naturally understand the so-called kinematic space~\cite{Czech:2015kbp,Czech:2015qta}, endow it with the interpretation of microscopic states such that to explain that the entropy is proportional to volume therein; in some sense quantitatively characterize the famous ``It from qubit'' thought experiment~\cite{VanRaamsdonk:2010pw}, that is, by removing the entanglement in the boundary quantum system, the bulk spacetime will accordingly deform, or even break up.

The idea of thread state will provide a complementary perspective distinct from the local tensors used in holographic TN models. Moreover, our thread/state prescription for locking thread configurations is an enlightening step towards the issue of spacetime emergence. It is intriguing to find the similar rules for the more general non-locking bit thread configurations, then it is possible to further read the SS states of the general bulk surfaces. It is even more tantalizing to completely reconstruct the bulk geometry merely from the properties of the quantum state assigned to the bit threads. In addition, it is also interesting to consider how the thread/state rules adapt to the covariant bit threads~\cite{Headrick:2022nbe} of the covariant RT formula~\cite{Hubeny:2007xt}, the quantum bit threads~\cite{Agon:2021tia,Rolph:2021hgz} that can account for the bulk quantum corrections to the RT formula~\cite{Faulkner:2013ana,Engelhardt:2014gca}, the Lorentzian bit threads~\cite{Pedraza:2021fgp,Pedraza:2021mkh} that can characterize the holographic complexity~\cite{Brown:2015bva,Susskind:2014rva}, the hyperthreads~\cite{Harper:2021uuq,Harper:2022sky} that can study the multipartite entanglement, and so on, and may provide useful insights on all of these topics.

\section{Prescription: thread/state correspondence} 
\begin{figure}[htbp]     \begin{center}
		\includegraphics[height=4.5cm,clip]{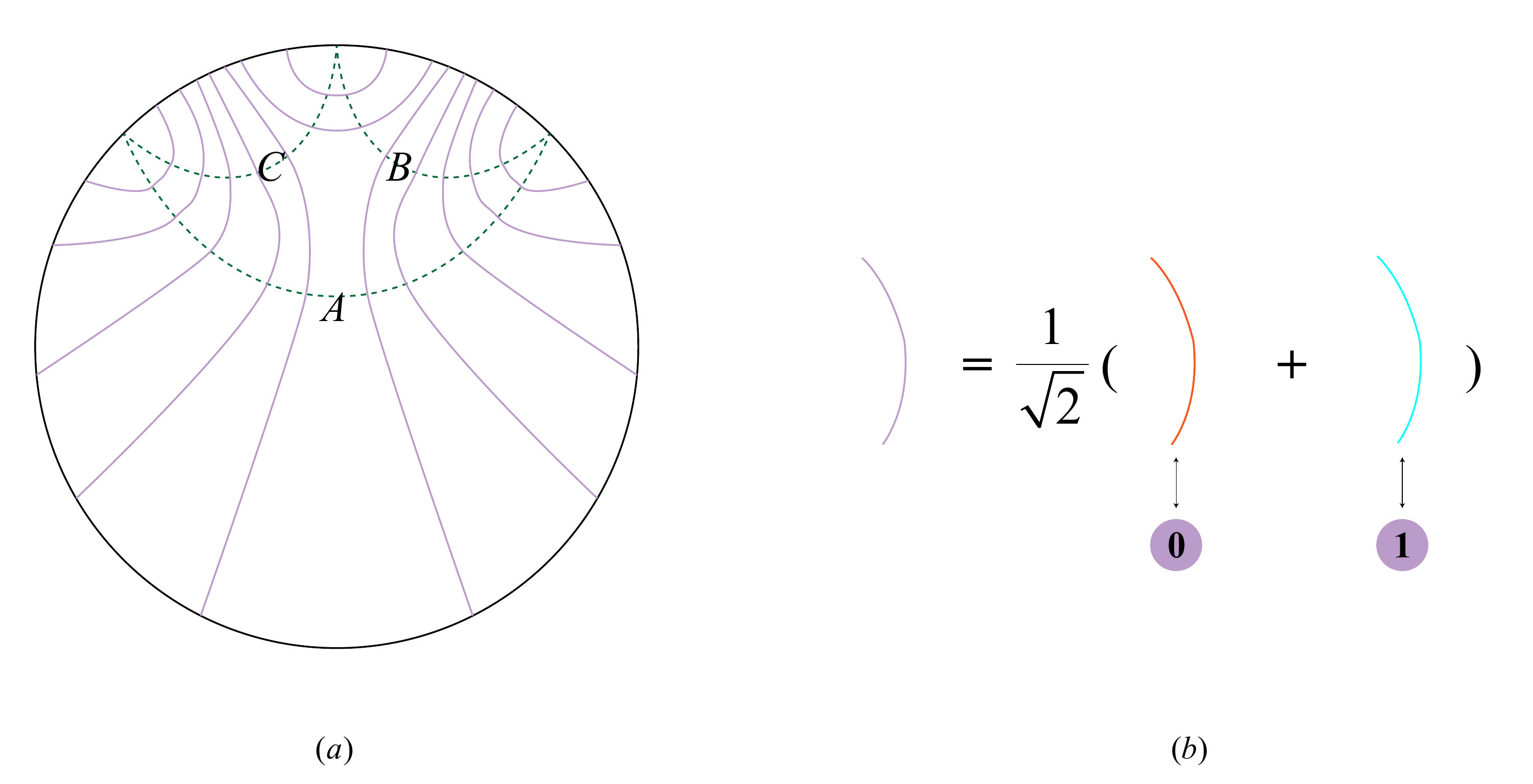}
		\caption{(a) A locking bit thread configuration that locks three RT surfaces $A$, $B$, and $C$ simultaneously. (b) Thread/state correspondence.	}
		\label{fig1}
	\end{center}	
\end{figure}
The motivation of thread/state correspondence originated from finding an effective scheme to characterize the entanglement structure between different bulk extremal surfaces in the framework of SS duality, or explicitly constructing the dual states (hereinafter referred to as SS states) of these extremal surfaces. This is inspired by the following two notable simple rules in the SS duality~\cite{Miyaji:2015fia,Miyaji:2015yva,Lin:2020ufd}: $\bf rule ~1:$ the density matrix corresponding to an extremal surface is a direct product of the density matrices at each point. In other words, a bulk extremal surface corresponds to an equal-probability mixed state; $\bf rule ~2:$ a closed surface corresponds to a pure state. Due to rule 1, one can imagine that on an extremal surface associated with an entropy $S$ (i.e., with an area of $4{G_N}S$), there distribute exactly $S$ bits (each with only two basic states, $\left| 1 \right\rangle $ and $\left| 0 \right\rangle $). Then, for any bulk extremal surface $\gamma$ with an entropy ${S_\gamma }$, we can label its basic states as $\left\{ {\left| {\gamma  = q} \right\rangle \left| {q \in N,0 \le q \le {2^{{S_\gamma }}} - 1} \right.} \right\}$. Each decimal number $q$ exactly corresponds to a binary string that describes the overall configuration state of all the SS bits on $\gamma$. The SS state corresponding to $\gamma$ can be represented as a mixed state ${\rho_\gamma }$ equipped with equal probabilities~\cite{Lin:2020ufd}:
\be\label{rho}{\rho _\gamma } = \sum\limits_{q = 0}^{{2^{{S_\gamma }}} - 1} {\frac{1}{{{2^{{S_\gamma }}}}}\left| {\gamma = q} \right\rangle \left\langle {\gamma = q} \right|} .\ee

Now we present the thread/state prescription. In the framework of the holographic principle, bit threads are defined as divergenceless but unoriented bulk threads whose density is less than $1/4{G_N}$ everywhere~\cite{Freedman:2016zud,Cui:2018dyq,Headrick:2017ucz}. Bit threads can be described by the language of $multiflow$. In particular, a bit thread configuration that can simultaneously optimize (i.e., maximize, or ``lock'') a set of thread fluxes through some specified boundary regions is referred to as a locking bit thread configuration~\cite{Headrick:2020gyq}. We will show that, following the prescription, a locking bit thread configuration can automatically determine the explicit form of the SS states of the bulk extremal surfaces. 

Consider a locking bit thread configuration that can lock three regions $A$, $B$, and $C$ (which are three RT extremal surfaces) simultaneously, as shown in figure~\ref{fig1}(a). Denote the numbers of threads connecting $AB$, $AC$, and $BC$ in the configuration as ${F_{AB}}$, ${F_{AC}}$, and ${F_{BC}}$ respectively, which are exactly half of the mutual information ${I_{AB}}$, ${I_{AC}}$, and ${I_{BC}}$~\cite{Lin:2020yzf}. The thread/state correspondence prescription is as follows: 

$(\bf 1)$ Each bit thread has two possible orthogonal states, namely $\left| {{\rm{red}}} \right\rangle $ state and $\left| {{\rm{blue}}} \right\rangle$ state. The same bit thread is always in a same ``color'' state, or a same superposition state of the color states, while the state of one bit thread does not affect the state of the other bit thread.

$(\bf 2)$ In a locking bit thread configuration, each bit thread is in a special quantum superposition state:
\be\label{thr}\left| {{\rm{thread}}} \right\rangle  = \frac{1}{{\sqrt 2 }}\left( {\left| {{\rm{red}}} \right\rangle  + \left| {{\rm{blue}}} \right\rangle } \right)\ee

$(\bf 3)$ In a locking bit thread configuration, the bit threads exactly meet the SS bits on the intersecting bulk extremal surfaces. The red state of a bit thread corresponds to the $\left| 0 \right\rangle$ state of the SS bits it passes through on the bulk extremal surfaces, while the blue state corresponds to the $\left| 1 \right\rangle $ state. In fact, this correspondence can be written more explicitly as
\be\begin{array}{l}
	\left| {{\rm{red}}} \right\rangle  = \left| {00 \cdots 00} \right\rangle \\
	\left| {{\rm{blue}}} \right\rangle  = \left| {11 \cdots 11} \right\rangle 
\end{array}\ee
In other words, if we perform a measurement operation to make a bit thread in a certain color state, the SS bits crossed by the same red bit thread will be all in the certain $\left| 0 \right\rangle$ state, while the SS bits crossed by the same blue bit thread will be all in the certain $\left| 1 \right\rangle$ state. 

Using the thread/state rules, we end up with an explicit form of the state of the whole closed surface $ABC$ as (see Appendix~\ref{appa} for details):
\be\label{full} \begin{array}{l}
	\left| \Psi  \right\rangle  = \sum\limits_{{q_{AB}} = 0}^{{2^{{F_{AB}}}} - 1} {\sum\limits_{{q_{AC}} = 0}^{{2^{{F_{AC}}}} - 1} {\sum\limits_{{q_{BC}} = 0}^{{2^{{F_{BC}}}} - 1} {\frac{1}{{\sqrt {{2^{{F_{AB}} + {F_{AC}} + {F_{BC}}}}} }}} } } \\
	\cdot \left( {\left| {{\phi _{AB}} = {q_{AB}}} \right\rangle  \otimes \left| {{\phi _{AC}} = {q_{AC}}} \right\rangle } \right)\\
	\otimes \left( {\left| {{\phi _{AB}} = {q_{AB}}} \right\rangle  \otimes \left| {{\phi _{BC}} = {q_{BC}}} \right\rangle } \right)\\
	\otimes \left( {\left| {{\phi _{AC}} = {q_{AC}}} \right\rangle  \otimes \left| {{\phi _{BC}} = {q_{BC}}} \right\rangle } \right)
\end{array}, \ee
where $\left| {{\phi _{AB}} = {q_{AB}}} \right\rangle  = \left| {\overbrace { -  -  \cdots  -  - }^{{F_{AB}}}} \right\rangle $ represents a configuration state of the bit threads connecting $A$ and $B$, whose number is ${F_{AB}}$, and it also provides the information of the configuration states of the SS bits on the bulk extremal surfaces that these bit threads pass through. The similar notation is used for ${\phi _{AC}}$ and ${\phi _{BC}}$. Note that (\ref{full}) is indeed a pure state, and the key point is that it is not difficult to verify that this expression (\ref{full}) can indeed be reduced to the correct reduced density matrixes (\ref{rho}) of $A$, $B$, and $C$.

A physical comment: because of the quantum entanglement between $B$ and $C$, the Hilbert space dimension of the states of $BC$ as a whole is exactly equal to the Hilbert space dimension of the corresponding states of $A$. Our thread/state correspondence rules explicitly show how the bit threads characterize the quantum entanglement between $B$ and $C$: that is, the SS bits on surfaces $B$ and $C$ crossed by the same bit thread must be in the same state. 

Moreover, it is immediately to apply this method to the case involving more RT surfaces. As long as the corresponding locking bit thread configuration is constructed, the SS states of these RT surfaces can be read out at the same time.

\section{Holographic qubit threads model} 
\begin{figure}[htbp]     \begin{center}
		\includegraphics[height=4cm,clip]{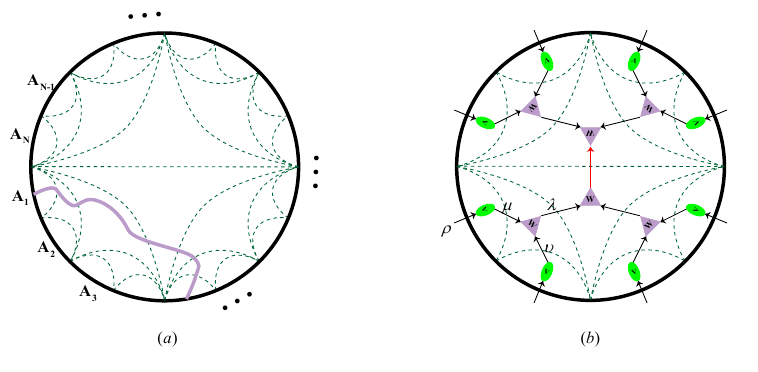}
		\caption{ (a) The holographic qubit threads model. We only present one component flow (the thick mauve line) schematically. The green dashed lines represent a set of non-intersecting RT surfaces. (b) Match with a holographic tensor network model.}
		\label{fig2}
	\end{center}	
\end{figure}
Inspired by the thread/state correspondence, it is natural to construct a novel toy model of the holographic principle, as shown in figure~\ref{fig2}(a). We can divide the holographic boundary quantum system into extremely large $N$ non-overlapping adjacent subregions (called $elementary ~regions$), denoted as ${A_1}$, ${A_2}$, ..., ${A_N}$, then introduce a $multiflow$ composed of $N\left( {N - 1} \right)/2$ $component~ flows$ to describe a locking thread configuration that can lock all these elementary regions and a set of non-intersecting $composite~ regions$ (the unions of elementary regions) simultaneously. In addition, we also require that each component flow flux in the locking bit thread configuration satisfies the so-called ``CFF=PEE'' scheme~\cite{Lin:2021hqs}, i.e.,
\begin{scriptsize}\be\label{fij}{F_{ij}} = \frac{1}{2}\left( {S\left( {\tilde A \cup {A_i}} \right) + S\left( {\tilde A \cup {A_j}} \right) - S\left( {\tilde A} \right) - S\left( {{A_i} \cup \tilde A \cup {A_j}} \right)} \right),\ee\end{scriptsize}
where we denote the region sandwiched between ${A_i}$ and ${A_j}$ as $\tilde A = {A_{\left( {i + 1} \right) \cdots \left( {j - 1} \right)}}$ .~\footnote{ See paper~\cite{Lin:2021hqs}, and it is not difficult to find that this expression is equivalent to the so-called PEE proposal proposed in~\cite{Kudler-Flam:2019oru,Wen:2018whg}. Moreover, in fact, PEE is referred to as conditional mutual information (CMI) in quantum information theory (up to a conventional 1/2 factor)~\cite{Czech:2015kbp,Czech:2015qta}, and (\ref{fij}) is just the definition of CMI.} This is for each component flow flux (CFF) to exactly match the partial entanglement entropy (PEE) in the boundary quantum system one by one, so as to reflect the entanglement structure information of the boundary quantum system correctly and reasonably. Using the convex dualization technique in the convex optimization theory and the bulk-cell decomposition method proposed in~\cite{Headrick:2020gyq}(see also~\cite{Lin:2020yzf}), we can prove that a locking thread configuration can lock a specified set of non-intersecting boundary regions simultaneously while subject to the extra constraints (\ref{fij}) at the same time always exists. See~\cite{app} for details.

According to the thread/state prescription, and following the previous notation, then one can write out the explicit density matrix of each RT surface directly. For a general connected boundary composite region $R = {A_{a\left( {a + 1} \right) \cdots b}} \equiv {A_a} \cup {A_{a + 1}} \cup  \cdots  \cup {A_b}$, the SS state of its corresponding holographic RT surface ${\gamma _{a\left( {a + 1} \right) \cdots b}}$ can be expressed as follows:
\be\begin{array}{l}
	{\rho _{{\gamma _{a\left( {a + 1} \right) \cdots b}}}} = \sum\limits_{s,t} {\sum\limits_{{q_{st}} = 0}^{{2^{{F_{st}}}} - 1} {\frac{1}{{{2^{{S_{a\left( {a + 1} \right) \cdots b}}}}}}} } \\
	\cdot \left( {\prod\limits_{s,t} {\left| {{\phi _{st}} = {q_{st}}} \right\rangle } } \right)\left( {\prod\limits_{s,t} {\left\langle {{\phi _{st}} = {q_{st}}} \right|} } \right)
\end{array}\ee

where $s \in \left\{ {a,a + 1, \cdots ,b} \right\}$, $t \notin \left\{ {a,a + 1, \cdots ,b} \right\}$.

\section{Relation   with   holographic   tensor   network   mo dels} 

Our holographic qubit threads model is closely related to the holographic tensor network models (the preliminary discussion on the connection between bit threads and holographic tensor networks can be seen in~\cite{Lin:2020yzf}). In fact, it is natural to convert the former form into the latter form. The key point is that a part of the whole thread configuration in each bulk ``cell'' divided by the RT surfaces can be written as a tensor using the thread/state prescription. In particular, these tensors can just be understood as the $disentanglers$ in the holographic tensor networks~\cite{Vidal:2007hda,Vidal:2008zz,Swingle:2009bg,Swingle:2012wq}.

More explicitly, as shown in figure~\ref{fig2}(b), one can construct two kinds of tensors using thread/state correspondence, denoted as $V$ and $W$. Wherein $V$ represents the distillation tensor~\cite{Bao:2018pvs,Bao:2019fpq,Lin:2020ufd}, which is defined as mapping the reduced state of each boundary elementary region to the SS state associated with its RT surface. In this paper, we do not focus on the study of $V$ tensor, which is merely a pro forma definition at the moment. The $W$ tensor plays the role of a $disentangler$~\cite{Vidal:2007hda,Vidal:2008zz,Swingle:2009bg,Swingle:2012wq}, which maps the SS states of the two RT surfaces in the previous layer to the SS state of the RT surface in the next layer.
Notice in the figure we also assign an arrow direction to each leg to specify the upper and lower indices of a tensor, then we can first write the $W$ tensor as $W_{\mu \nu }^\lambda $ pro forma, where $\mu $ and $\nu $ represent the SS states of the two RT surfaces of the previous layer, respectively, while $\lambda $ represents the SS state of the RT surface of the inner layer. Then the thread/state correspondence indicates that $\mu $, $\nu $ and $\lambda $ could be further decomposed into smaller indices, thus one can write a universal expression for the $W$ tensor of any layer immediately. We first write down the corresponding pure state of the closed surface that is the boundary of the cell of each layer. Consider a cell consisting of three RT surfaces ${\gamma _R}$, ${\gamma _S}$ and ${\gamma _T}$ corresponding to three adjacent composite boundary regions $R$, $S$ and $T$, respectively. Let us denote $R =  \cup {A_r}$, $S =  \cup {A_s}$, $T =  \cup {A_t}$, i.e., denote the elementary regions within $R$, $S$ and $T$ as ${A_r}$, ${A_s}$ and ${A_t}$ respectively. Following these notations, we can write:
\begin{small}\be \begin{array}{l}
		\left| {{\Psi _{{\gamma _R} \cup {\gamma _S} \cup {\gamma _T}}}} \right\rangle  = \left( {\sum\limits_{r,t} {\sum\limits_{{q_{rt}} = 0}^{{2^{{F_{rt}}}} - 1} {} } } \right)\left( {\sum\limits_{s,t} {\sum\limits_{{q_{st}} = 0}^{{2^{{F_{st}}}} - 1} {} } } \right)\left( {\sum\limits_{r,s} {\sum\limits_{{q_{rs}} = 0}^{{2^{{F_{rs}}}} - 1} {} } } \right)\\
		\frac{1}{{\sqrt {{2^{\sum\limits_{r,t} {{F_{rt}}}  + \sum\limits_{s,t} {{F_{st}}}  + \sum\limits_{r,s} {{F_{rs}}} }}} }} \cdot \left( {\prod\limits_{r,t} {\left| {{\phi _{rt}} = {q_{rt}}} \right\rangle } } \right)\\
		\otimes \left( {\prod\limits_{s,t} {\left| {{\phi _{st}} = {q_{st}}} \right\rangle } } \right) \otimes \left( {\prod\limits_{r,s} {\left| {{\phi _{rs}} = {q_{rs}}} \right\rangle } } \right)
	\end{array},\ee\end{small}

then we obtain
\be\label{w} \begin{array}{l}
	W_{{\mu _R}{\nu _S}}^{{\lambda _T}} = W_{\left\{ {{\mu _{rt}}} \right\}\left\{ {{\mu _{rs}}} \right\}\left\{ {{\nu _{rs}}} \right\}\left\{ {{\nu _{st}}} \right\}}^{\left\{ {{\lambda _{rt}}} \right\}\left\{ {{\lambda _{st}}} \right\}}\\
	\quad \quad \, = \left\{ {\begin{array}{*{20}{c}}
			{\frac{1}{{\sqrt {{2^{\sum\limits_{r,t} {{F_{rt}}}  + \sum\limits_{s,t} {{F_{st}}}  + \sum\limits_{r,s} {{F_{rs}}} }}} }},\quad {\lambda _{ij}} = {\mu _{ij}} = {\nu _{ij}}}\\
			{0.\quad otherwise}
	\end{array}} \right.
\end{array}.\ee

From the expression of (\ref{w}), it is clear that the lower index of the $W$ tensor has the $rs$ component representing the entanglement between a pair of elementary regions ${A_r}$ and ${A_s}$, while the upper index has no $rs$ component. Therefore, the $W$ tensor is exactly playing the role of a $disentangler$ (of course, it also plays the role of the $coarse$-$grainer$ at the same time. A preliminary discussion of this issue can be seen in~\cite{Lin:2020ufd,Lin:2020yzf}), because it is such a unitary operation that just converts the state of two entangled blocks in the previous layer into a state of a block without internal entanglement. This process of disentanglement can be intuitively understood as removing the contribution of the inner threads that represent the shorter range entanglement! This process is carried out successively, exactly in line with the idea of MERA-like tensor network, that is, the effect of short-range entanglement is successively removed and only the effect of long-range entanglement is finally focused on~\cite{Vidal:2007hda,Vidal:2008zz,Swingle:2009bg,Swingle:2012wq}.

On the other hand, we can directly write the distillation tensor $V$ pro forma as
\be V_{{\rho _i}}^{{\mu _i}} = V_{{\rho _i}}^{\left\{ {{\mu _{ij}}} \right\}},\ee
where $j$ runs from 1 to $N$ (except for $j$ itself), and the lower index ${\rho _i}$ represents the reduced state of the elementary region ${A_i}$, while ${\mu _i}$ represents the SS state of the RT surface ${\gamma _i}$ corresponding to ${A_i}$. For example, for region ${A_1}$, 
\be V_{{\rho _1}}^{{\mu _1}} = V_{{\rho _1}}^{{\mu _{12}}{\mu _{13}} \cdots {\mu _{1N}}}.\ee
The $V$ tensors are isometries~\cite{Bao:2018pvs,Bao:2019fpq,Lin:2020ufd}.

\section{Relation with kinematic space } 
Our holographic qubit threads model also has a close connection to the so-called kinematic space~\cite{app}. Kinematic space is an example of the application of quantum information theory to the holographic duality~\cite{Czech:2015kbp,Czech:2015qta}. Briefly speaking, it is a dual space wherein each point is one-to-one mapped from a pair of points on the boundary of the original geometry.~\cite{Czech:2015kbp,Czech:2015qta} proposed that in the framework of the holographic principle, the metric of a kinematic space (i.e., its spatial volume density) is defined by the conditional mutual information (CMI),
\be\label{met} \mathscr{D} g  \equiv {d^2}{V_{{\rm{kin}}}} = {\rm{CMI}},\ee
where
\begin{small}
	\be{\rm{CMI(}}A,C\left| B \right.{\rm{)}} = S(AB) + S(BC) - S(ABC) - S(B),\ee
\end{small}
and the entanglement entropy can be represented by a volume in kinematic space
\be\label{vol} S \propto {\rm{Vol}}.\ee
Naturally, we can relate each thread in our holographic qubit threads model to a point in a kinematic space. Moreover, seeing (\ref{fij}), CMI and PEE (or our component flow flux, CFF) are exactly the same thing characterizing the entanglement density in a sense. The point is that according to our thread/state correspondence, now every point in kinematic space can be just interpreted as a qubit, namely a quantum superposition of two orthogonal basic states. Therefore, in fact, our thread/state rules can be regarded as indicating the microstates of the kinematic space. By thread/state rules, 
\be S \propto \ln \Omega  = \ln \left( {{2^{\sum {{F_{ij}}} }}} \right) = \sum {{F_{ij}}} ,\ee
by (\ref{fij}),
\be\sum {{F_{ij}}}  = \sum {{\rm{CMI}}} ,\ee 
then by (\ref{met}), one obtain (\ref{vol}).

In other words, simply counting the number of these microstates $\Omega$ leads to the conclusion that the entanglement entropy is proportional to the volume of kinematic space (a natural interpretation). Kinematic space is also closely related to the interpretation of the MERA tensor network~\cite{Czech:2015kbp,Czech:2015qta}, therefore, we expect that our thread/state correspondence will deepen the understanding of both.

\section{Relation with the connectivity of spacetime} 

\begin{figure}[htbp]     \begin{center}
		\includegraphics[height=6cm,clip]{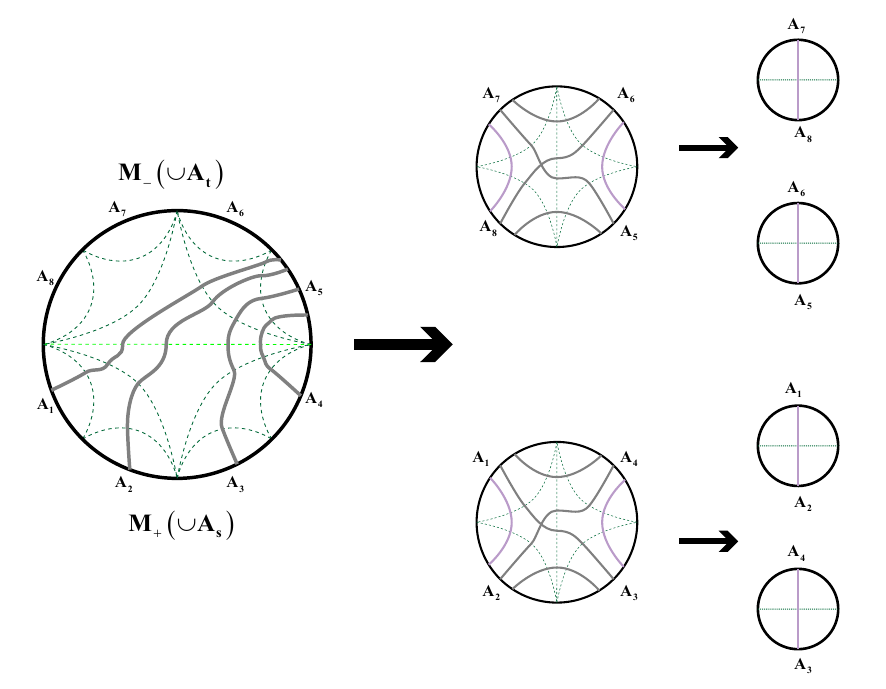}
		\caption{Removing all entanglement between the set of ${A_s}$ and the set of ${A_t}$ (we only draw a few of relevant thread bundles, i.e., the thick grey lines in the figure), the original bulk spacetime $N$ is split into two independent holographic bulk spacetimes ${N_ + }$ and ${N_ - }$. With this process repeating, finally, a bulk spacetime $N$ will eventually disintegrate into a tremendous number of small bubbles.}
		\label{fig9}
	\end{center}	
\end{figure}

Our holographic qubit threads model can visualize the connection between the entanglement of the boundary quantum system and the connectivity of the holographic bulk spacetime. Consider a thought experiment similar to Raamsdonk's famous ``It from qubit" experiment~\cite{VanRaamsdonk:2010pw}, see figure~\ref{fig9}. Consider dividing the whole boundary quantum system $M$ into two halves, denoted as ${M_ + }$ and ${M_ - }$, then remove the entanglement between the two halves, but still preserve the entanglement between the internal elementary regions within ${M_ + }$ and those within ${M_ - }$. Then this implies that the values of bit thread fluxes characterizing the former vanish, while those characterizing the latter are unchanged. Since these vanishing bit threads are originally passing through a series of RT surfaces and thus contributing to all these RT surfaces, not only does the area of the RT surface separating ${M_ + }$ and ${M_ - }$ shrink to zero (and thus the original bulk spacetime $N$ is split into two new disconnected independent spacetimes), but the areas of the RT surfaces inside the two halves of $N$ also decreases. Although we still do not fully figure out how to decode the geometric information of any surface or any region in the bulk from the entanglement information in the boundary CFT states completely, however, the area changes of these RT surfaces can quantitatively characterize how the bulk spacetime shrinks and splits as the quantum entanglement is taken away to some extent. This quantitative calculation can in principle be explicitly implemented by the following formula in the locking bit thread configuration~\cite{Lin:2021hqs}:
\be\begin{array}{l}
	{\rm{Are}}{{\rm{a}}_{a\left( {a + 1} \right) \ldots b}} = 4{G_N}\sum\limits_{s,t} {{F_{st}}} ,\\
	s \in \left\{ {a,a + 1, \cdots ,b} \right\},t \notin \left\{ {a,a + 1, \cdots ,b} \right\}
\end{array},\ee
where ${\rm{Are}}{{\rm{a}}_{a\left( {a + 1} \right) \ldots b}}$ represents the area of the RT surface associated with a connected composite region $R = {A_{a\left( {a + 1} \right) \cdots b}} \equiv {A_a} \cup {A_{a + 1}} \cup  \cdots  \cup {A_b}$.
This is very interesting. With this splitting process repeating, finally, a bulk spacetime $N$ will eventually disintegrate into a tremendous number of small bubbles. Conversely, this implies that, one can build a continuous spacetime using quantum entanglement. Figuratively speaking, bit threads, or we now should say, ``qubit threads'' can play the roles of ``sewing'' a spacetime. The threads extract the entanglement information from the boundary quantum system, and then build a spacetime just like sewing many small fragments into a sweater!

\section{Discussions}\label{sect:conclusion}

The idea of holographic tensor networks or quantum information theory has proved to be enlightening to the issue of spacetime emergence (``it from qubit''), while in this letter we have seen that the concept of (long-range) ``threads'' can provide a new perspective that is different from, but closely related to the local ``tensors'' (or quantum circuit gates). Although our thread/state correspondence rules for locking thread configurations merely apply to bulk extremal surfaces at present, it is a suggestive step towards the issue of spacetime emergence. Since the concept of bit threads originates from the duality of the optimization problem of the areas of geometric surfaces and the optimization problem of the fluxes of thread flows, we can further ask whether similar rules can be found and applied to the more general non-locking bit thread configurations so that one can further read the SS states of the more general bulk surfaces. In the longer term, an even more tantalizing idea is to further cast off the metric information of the background spacetime and completely reconstruct the bulk geometry only from the properties of the quantum state assigned to the bit threads.

On the other hand, actually there are various updated versions of bit threads, which are very useful tools for studying various aspects of the relationship between geometry and quantum entanglement, such as the covariant bit threads~\cite{Headrick:2022nbe}, the quantum bit threads~\cite{Agon:2021tia,Rolph:2021hgz}, the Lorentzian bit threads~\cite{Pedraza:2021fgp,Pedraza:2021mkh}, the hyperthreads~\cite{Harper:2021uuq,Harper:2022sky}, etc. It is likely that our thread/state correspondence rules could be further adapt to these similar objects and lead to deeper or clearer understandings of these different important aspects of the holographic duality.

\section*{Acknowledgement}
We would like to thank Ling-Yan Hung and Sun Yuan for useful discussions.

\begin{appendix}

\section{A user guide to thread/state prescription}\label{appa}
Considering the simplest case involving three RT surfaces ($A$, $B$, and $C$) as shown in figure~\ref{fig1}(a), focusing on the correct equal-probability mixed states ${\rho _A}$, the pure state corresponding to the whole closed surface $ABC$ can be constructed as
\be\label{pure}\left| \Psi  \right\rangle  = \sum\limits_{q = 0}^{{2^{{S_A}}} - 1} {\frac{1}{{\sqrt {{2^{{S_A}}}} }}\left| {A = q} \right\rangle  \otimes \left| {BC = q} \right\rangle } ,\ee
where $\left| {BC = q} \right\rangle $ is merely a pro forma definition at present, representing the basic states of the complement of region $A$ in the whole closed system. With a little thought, we realize that each of these $\left| {BC = q} \right\rangle$ states should actually be a quantum superposition state of $\left| {\underbrace { -  -  \cdots  -  - }_{{S_B}}\underbrace { -  -  \cdots  -  - }_{{S_C}}} \right\rangle  = \left| {\underbrace { -  -  \cdots  -  - }_{{S_B}}} \right\rangle  \otimes \left| {\underbrace { -  -  \cdots  -  - }_{{S_C}}} \right\rangle  \equiv \left| {B = {q_b}} \right\rangle  \otimes \left| {C = {q_c}} \right\rangle$.

Now suppose we are going to measure surface $A$ and find that $A$ is in a certain configuration after the measurement operation, for example, in the state $\left| {A = 0} \right\rangle  \equiv \left| {\underbrace {00 \cdots 00}_{{S_A}}} \right\rangle $. Then by {\bf rule 3}, this is also equivalent to measuring that all bit threads passing through surface $A$ (i.e., starting from $A$, and connecting to $B$ or $C$) are in red state. As per {\bf rule 1}, the operation of measuring $A$ will not affect the state of other bit threads that do not cross $A$ (i.e. threads connecting $B$ with $C$), then we can immediately construct the explicit form of the corresponding configuration$\left| {BC = 0} \right\rangle$ that $BC$ must be in when $A$ is measured to be  in configuration $\left| {A = 0} \right\rangle  \equiv \left| {\underbrace {00 \cdots 00}_{{S_A}}} \right\rangle $. Firstly, under this measurement, that $A$ is in configuration $\left| {A = 0} \right\rangle  \equiv \left| {\underbrace {00 \cdots 00}_{{S_A}}} \right\rangle$ corresponds to the bit threads passing through surface $A$ (to $B$ or $C$) are all in red state, therefore, in fact, we have determined the states that a certain part of the SS bits on surface $B$ and $C$ must be in, that is, the SS bits passed through by these red bit threads are all in the definite state $\left| 0 \right\rangle $. On the other hand, the other bits on $B$ and $C$ are passed through by the bit threads in the superposition state (\ref{thr}) ({\bf rule 2}). Therefore, the prescription automatically produces the specific form of $\left| {BC = 0} \right\rangle $ as
\be\begin{array}{l}
	\left| {BC = 0} \right\rangle  = \chi  \cdot \{ \left| {\underbrace {\overbrace {00 \cdots 00}^{{F_{AB}}}\overbrace {00 \cdots 00}^{{F_{BC}}}}_{{S_B}}\underbrace {\overbrace {00 \cdots 00}^{{F_{AC}}}\overbrace {00 \cdots 00}^{{F_{BC}}}}_{{S_C}}} \right\rangle \\
	+ \left| {\underbrace {\overbrace {00 \cdots 00}^{{F_{AB}}}\overbrace {00 \cdots 01}^{{F_{BC}}}}_{{S_B}}\underbrace {\overbrace {00 \cdots 00}^{{F_{AC}}}\overbrace {00 \cdots 01}^{{F_{BC}}}}_{{S_C}}} \right\rangle \\
	+  \cdots  + \left| {\underbrace {\overbrace {00 \cdots 00}^{{F_{AB}}}\overbrace {11 \cdots 11}^{{F_{BC}}}}_{{S_B}}\underbrace {\overbrace {00 \cdots 00}^{{F_{AC}}}\overbrace {11 \cdots 11}^{{F_{BC}}}}_{{S_C}}} \right\rangle \} 
\end{array},\ee

	where $\chi $ is a normalization constant that should satisfy
	\be{\chi ^2} \cdot {2^{{F_{BC}}}} = 1,\ee
	that is,
	\be\chi  = \frac{1}{{\sqrt {{2^{{F_{BC}}}}} }} = \frac{1}{{\sqrt {{2^{\frac{{{S_B} + {S_C} - {S_A}}}{2}}}} }}.\ee
	
	Similarly, we can use the same reasoning to obtain the explicit form of the corresponding $\left| {BC = q} \right\rangle $ state for each $\left| {A = q} \right\rangle $ state, and finally obtain the expression of the full pure state of the whole system $ABC$ as follows:
\begin{tiny}\be\label{full2}\begin{array}{l}
		\left| \Psi  \right\rangle  =\\
		 \frac{1}{{\sqrt {{2^{{F_{AB}} + {F_{AC}}}}} }}\left| {\underbrace {\overbrace {00 \cdots 00}^{{F_{AB}}}\overbrace {00 \cdots 00}^{{F_{AC}}}}_{{S_A}}} \right\rangle  \otimes \{ \frac{1}{{\sqrt {{2^{{F_{BC}}}}} }}(\left| {\underbrace {\overbrace {00 \cdots 00}^{{F_{AB}}}\overbrace {00 \cdots 00}^{{F_{BC}}}}_{{S_B}}\underbrace {\overbrace {00 \cdots 00}^{{F_{AC}}}\overbrace {00 \cdots 00}^{{F_{BC}}}}_{{S_C}}} \right\rangle \\
		+ \left| {\underbrace {\overbrace {00 \cdots 00}^{{F_{AB}}}\overbrace {00 \cdots 01}^{{F_{BC}}}}_{{S_B}}\underbrace {\overbrace {00 \cdots 00}^{{F_{AC}}}\overbrace {00 \cdots 01}^{{F_{BC}}}}_{{S_C}}} \right\rangle  +  \cdots  + \left| {\underbrace {\overbrace {00 \cdots 00}^{{F_{AB}}}\overbrace {11 \cdots 11}^{{F_{BC}}}}_{{S_B}}\underbrace {\overbrace {00 \cdots 00}^{{F_{AC}}}\overbrace {11 \cdots 11}^{{F_{BC}}}}_{{S_C}}} \right\rangle )\} \\
		+ \frac{1}{{\sqrt {{2^{{F_{AB}} + {F_{AC}}}}} }}\left| {\underbrace {\overbrace {00 \cdots 00}^{{F_{AB}}}\overbrace {00 \cdots 01}^{{F_{AC}}}}_{{S_A}}} \right\rangle  \otimes \{ \frac{1}{{\sqrt {{2^{{F_{BC}}}}} }}(\left| {\underbrace {\overbrace {00 \cdots 00}^{{F_{AB}}}\overbrace {00 \cdots 00}^{{F_{BC}}}}_{{S_B}}\underbrace {\overbrace {00 \cdots 01}^{{F_{AC}}}\overbrace {00 \cdots 00}^{{F_{BC}}}}_{{S_C}}} \right\rangle \\
		+ \left| {\underbrace {\overbrace {00 \cdots 00}^{{F_{AB}}}\overbrace {00 \cdots 01}^{{F_{BC}}}}_{{S_B}}\underbrace {\overbrace {00 \cdots 01}^{{F_{AC}}}\overbrace {00 \cdots 01}^{{F_{BC}}}}_{{S_C}}} \right\rangle  +  \cdots  + \left| {\underbrace {\overbrace {00 \cdots 00}^{{F_{AB}}}\overbrace {11 \cdots 11}^{{F_{BC}}}}_{{S_B}}\underbrace {\overbrace {00 \cdots 01}^{{F_{AC}}}\overbrace {11 \cdots 11}^{{F_{BC}}}}_{{S_C}}} \right\rangle )\} \\
		+  \cdots \\
		\frac{1}{{\sqrt {{2^{{F_{AB}} + {F_{AC}}}}} }}\left| {\underbrace {\overbrace {11 \cdots 11}^{{F_{AB}}}\overbrace {11 \cdots 11}^{{F_{AC}}}}_{{S_A}}} \right\rangle  \otimes \{ \frac{1}{{\sqrt {{2^{{F_{BC}}}}} }}(\left| {\underbrace {\overbrace {11 \cdots 11}^{{F_{AB}}}\overbrace {00 \cdots 00}^{{F_{BC}}}}_{{S_B}}\underbrace {\overbrace {11 \cdots 11}^{{F_{AC}}}\overbrace {00 \cdots 00}^{{F_{BC}}}}_{{S_C}}} \right\rangle \\
		+ \left| {\underbrace {\overbrace {11 \cdots 11}^{{F_{AB}}}\overbrace {00 \cdots 01}^{{F_{BC}}}}_{{S_B}}\underbrace {\overbrace {11 \cdots 11}^{{F_{AC}}}\overbrace {00 \cdots 01}^{{F_{BC}}}}_{{S_C}}} \right\rangle  +  \cdots  + \left| {\underbrace {\overbrace {11 \cdots 11}^{{F_{AB}}}\overbrace {11 \cdots 11}^{{F_{BC}}}}_{{S_B}}\underbrace {\overbrace {11 \cdots 11}^{{F_{AC}}}\overbrace {11 \cdots 11}^{{F_{BC}}}}_{{S_C}}} \right\rangle )\} 
	\end{array}.\ee\end{tiny}	
Notice that therein ${F_{AB}} + {F_{AC}} = {S_A}$. Rewriting (\ref{full2}) compactly then results  in the simple and symmetric form (\ref{full}).
	
\vspace{5cm}

\end{appendix}




\end{document}